Embracing Sex-specific Differences in Engineered Kidney Models for Enhanced Biological Understanding

Charlotte A Veser[1], Aurélie MF Carlier[2], Silvia M Mihăilă[1*], Sangita Swapnasrita[2*]

[1] Utrecht Institute for Pharmaceutical Sciences, Universiteitsweg 99, 3584 CG Utrecht, the Netherlands

[2] MERLN Institute for Technology-Inspired Regenerative Medicine, Universiteitssingel 40, 6229 ER Maastricht, the Netherlands

* *Both authors contributed equally*
Email: s.swapnasrita@maastrichtuniversity.nl, s.mihaila@uu.nl

**Abstract**

*In vitro* models play a crucial role in advancing our understanding of biological processes, disease mechanisms, and developing screening platforms for drug discovery. Kidneys play an instrumental role in transport and elimination of drugs and toxins. However, despite the well-established patient-to-patient differences in kidney function and disease manifestation, progression and prognostic, few studies take this variability into consideration. In particular, the discrepancies between female and male biology warrants a better representation within kidney *in vitro* models. The omission of sex as a biological variable poses the significant risk of overlooking sex-specific mechanisms in health and disease and potential differences in drug efficacy and toxicity between males and females. This review aims to highlight the importance of incorporating sex dimorphism in kidney *in vitro* models by examining the sexual characteristics in the context of the current state-of-the-art. Furthermore, this review underscores opportunities for improving kidney models by incorporating sex-specific traits. Ultimately, this roadmap to incorporating sex-dimorphism in kidney *in vitro* models will facilitate the creation of better models for studying sex-specific mechanisms in the kidney and their impact on drug efficacy and safety.

The intersection of tissue engineering and patient-specific stem cell biology has ushered in an era of innovative in vitro tissue models, opening avenues for tailored treatment approaches. Yet, amid these advancements, the crucial factor of gender-based disparities in health and disease has often been overlooked within the realms of stem cell biology, tissue engineering, and preclinical screening. These disparities, spanning from systemic hormonal influences to nuanced cellular-level variations, hold profound implications that deserve closer attention. Significantly, the renal system exhibits noteworthy gender-based distinctions that warrant recognition and integration into kidney tissue engineering endeavors. In this extensive review, we delve into gender-specific properties of kidney tissue, considering both health and disease contexts. Furthermore, we propose a framework for incorporating gender-based differences into the realm of human kidney tissue engineering. This review elucidates the avenues through which gender-based characteristics can be harnessed at the cellular and tissue levels. By contemplating the development of gender-specific kidney models, we offer a promising perspective on advancing research into renal diseases. Additionally, we outline essential design criteria to support the growth of sex-specific kidney tissue engineering. In conclusion, this review not only underscores the significance of acknowledging gender-related disparities in tissue engineering but also sets the stage for future research possibilities beyond the renal system, paving the way for a more comprehensive understanding of patient-specific treatment modalities in the context of kidney tissue engineering.

**Highlights**

- Sex-specific differences in kidney function, especially in solute and drug handling are often under reported.
- The impact of sex hormones on kidney physiology and pathology is overlooked.
- A practical guide for researchers to incorporate sex as a biological variable and improve reporting standards is proposed.

**Background**

Understanding sex-specific differences in how kidneys function and respond to diseases is crucial for developing safer and more effective drugs. However, most *in vitro* representations of the kidney fail to capture these characteristics, limiting our understanding of kidney health and jeopardizing efficient pharmacotherapy. Herein, we explore the current knowledge on sex differences in kidney function and disease, as well as the limitations of existing kidney *in vitro* models. By addressing these limitations, we can create improved models that will not only enhance our understanding of kidney diseases, but also lead to better drug discovery and safety assessments. This means more cost effective and safer drug development processes, benefiting both patients and clinical trials. By leveraging the power of *in vitro* models and considering sex differences, we can unlock new insights into kidney health, improve drug development, and pave the way for personalized and effective treatments for everyone.

**Keywords**: kidney, sex-dimorphism, sex-specificity, *in vitro* models, nephrotoxicity

# 1 Introduction

The kidney is a vital organ responsible for maintaining proper blood pressure, removing of metabolic waste and regulating fluid and electrolyte levels. Kidney diseases are a significant global health concern, and they exhibit substantial disparities associated with sex. These differences, rooted in biology and physiology, create a diverse landscape of kidney health and disease experiences between men and women. Men and women exhibit disparities not only in their susceptibility to kidney disease and risk factors but also in various biological processes, such as aging, cell apoptosis, and the operation of key homeostatic systems like blood pressure, fluid balance, and the hypothalamic-pituitaryadrenal axis. These differences in disease progression can be attributed to several factors, including the presence or absence of a Y chromosome, variations in gene expression, inheritance patterns of the mitochondrial genome, and differences in neurohormonal activity (1).

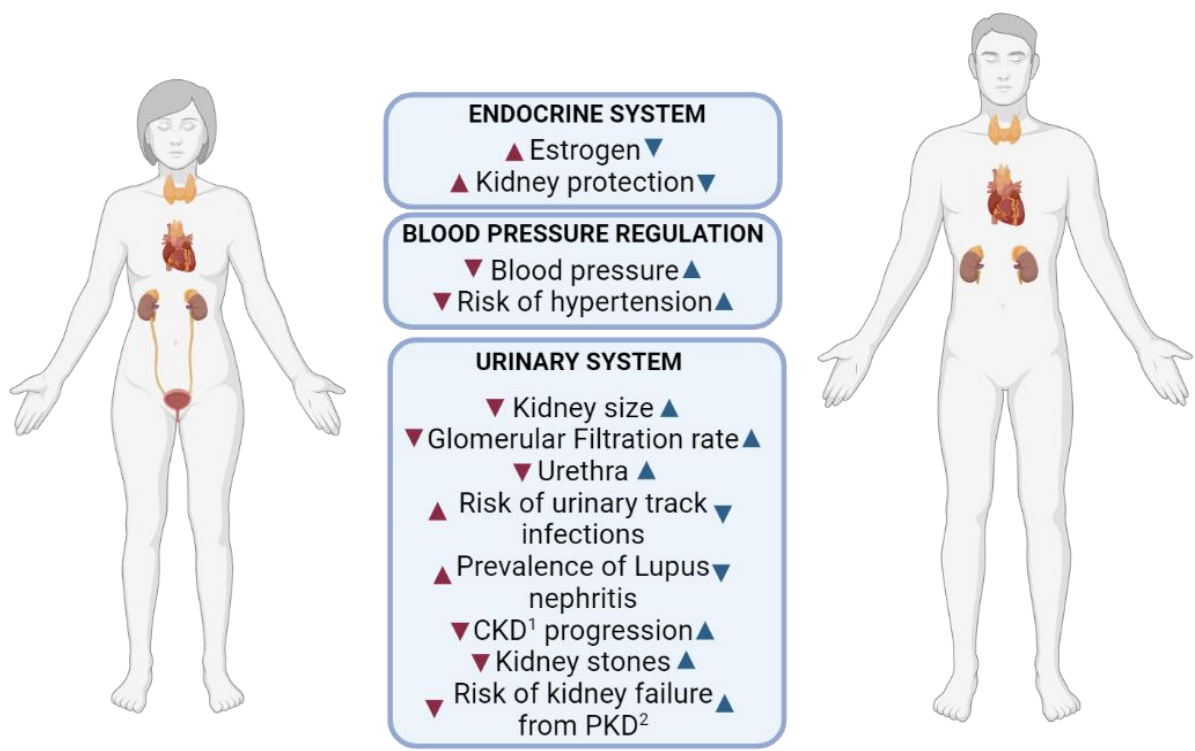

**Figure 1**: Known differences in kidney physiology, hormonal regulation and prevalence of kidney related disorders between the sexes. [1] chronic kidney disease, [2] polycystic kidney disease.

However, like in cardiovascular research, sex is often an overlooked variable in preclinical and clinical studies related to kidney health and disease. Historically, clinical and preclinical trials have predominantly involved male participants, driven by the assumption that sex-related differences primarily pertain to reproductive biology and that hormonal fluctuation in females are causing high experimental variability, despite the FDA mandate requesting the inclusion of females in clinical trials since 1993. However, the evolving perspective recognizes the importance of considering sex differences in research settings more comprehensively. Various initiatives have been created to close this gap. In 1999, the Committee on Understanding the Biology of Sex and Gender Differences formed by the Institute of Medicine, USA, recommended expanding the "research on sex at a cellular level" from birth to death and across species in an attempt to eliminate "barriers to advancement of science in health and illness" (2). Nonetheless, emerging evidence has demonstrated that sex differences in kidney function and disease do exist, highlighting the need to incorporate sex as an important variable in research(3). The National Institutes of Health (NIH) has since 2015 mandated that any NIH-funded research considers sex as a biological variable and should therefore be factored into the design, analysis, and reporting (4). Following the call by NIH, many journals have amended their policies to include sex-specific reporting (5, 6). As a part of the Gender Equality Strategy 2020-2025, Horizon Europe defines the integration of sex as a research variable by default and to be evaluated under the excellence criteria (7). Other global granting agencies have also recently followed a similar inclusive policy towards sex-specific representation in research (8).

The observed clinical disparities observed in kidney health diseases arise from the complex interplay between biological sex and socially constructed gender (see section 2). Common disparities manifest in various aspects of kidney disorders, including physiology, hormonal regulation and disease prevalence are shown in **Figure 1** (9-16). Furthermore, how individuals respond to medications can be influenced by sex-related differences in drug metabolism and response. This factor underscores the importance of tailoring medications and treatment strategies based on a patient's sex and investigating the intricate relationship between sex and kidney (patho)physiology holds considerable promise. Although sex-specific hormones play a significant role in kidney function, they are not the only factor. Gender-specific stressors, such as perceived stress, was associated with poorer quality of life in men (17). The same authors showed that women have cultivated coping strategies which helps them to maintain a higher quality of life. Moreover, recent evidence suggests that the impact of sex hormones on kidney function may vary depending on the underlying disease and stage of the disease. Therefore, understanding the role of sex in kidney disease requires a comprehensive investigation of the multiple sex-related factors involved. Such investigations have the potential to deepen our understanding of the mechanisms underlying kidney diseases, enhance the safety and effectiveness of drug development, and enable the development of more targeted interventions that cater to individuals of all genders.

Although kidney transport machinery are poorly studied, many physiological differences have been observed in male and female kidney (section 3). *In vitro* kidney models, supported by stem cell biology, tissue engineering and organ on chip manufacturing, offer unique opportunities to study sex differences in kidney function and disease (section 4). The incorporation of sex-specificity into *in vitro* kidney models can provide insight into the mechanisms contributing to sexual dimorphisms in humans (section 5). By using these models, we can parse out the individual and synergistic roles of various sex-related factors in regulating kidney function, which may ultimately improve our understanding of the role of sex in health and disease.

## 2   Why do we see sex dimorphism in humans?

To discuss sex dimorphism in a comprehensive way, it is first necessary to understand the distinction between sex and gender. **Sex** refers to biologically determined characteristics, such as chromosomes, sex organds, and endogeneous hormone profiles, (18, 19) while 'gender' encompasses socially constructed attributes (20, 21). Importantly, neither sex nor gender conforms strictly to a binary model, and various sexes exist (for example, Turner syndrome (X), Klinefelter syndrome (XXY) and XYY or XXXY syndromes (22)).  In this review, we primarily focus on comparing the biological sexes, namely male (XY) and female (XX), which are referred to as men and women, respectively.

**Table 1:** Contributors to sex dimorphism in humans with XX and XY sex chromosomes including examples of kidney physiology and pathology.

| XX | XY |
|---|---|
| **Sex chromosomes(20, 23):** | |
| - Two X chromosomes<br>- More severe effects in X-linked recessive disorders such as X-linked dominant hypophosphatemic rickets and Fabry disease(23). | - One X and one Y chromosome<br>- SRY gene on Y chromosome regulates testis development |
| **X chromosome inactivation(23-25):**<br>Silencing of one chromosome from gene expression through histone modification and DNA methylation to balance X- linked gene dosage between sexes. | |
| - The inactivated X chromosome is not identical in all cells (mosaicism)<br>- Distribution of X inactivation is unequal or shifts over time (skewing) | - Ubiquitous Y gene expression<br>- No need for X chromosome inactivation, but a more severe manifestation of Alport Syndrome, if present. |
| • Genes escape inactivation. Mutation/alterations may lead to kidney disorders such as Alport Syndrome(23, 25, 26), X-linked nephropathy(23), etc. | |
| **Imprinting (24, 27):** Differential gene expression is dependent on the parent the gene was inherited from. Abnormal imprinting can lead to an increased risk of Wilms tumor and autosomal dominant polycystic kidney disease (sex-independent in rats(28)). | |
| • Both maternal and paternal imprinting of X chromosomes | • Only maternal X chromosome imprinting |
| **Sex steroids(29):** Sex steroid hormones such as estradiol, testosterone, progesterone, luteinizing hormone (LH), and follicular stimulating hormone (FSH) are present in both sexes. Sex steroids influence a variety of signalling pathways, permanently changing cell epigenetics. | |
| • High estradiol levels after puberty until levels drop drastically at menopause. It is generally accepted that estrogen has a protective effect on kidneys(30-32). | • High testosterone peaks at puberty and remains stable with a continuous decline |
| **Lifestyle and aging(20, 33, 34):** | |
| • Differences in lifestyle accumulate over lifetime as differences in epigenetics, DNA methylation, histone modification, chromatin architecture, and miRNA expression. For example, a high sodium diet could lead to an increased risk of chronic kidney disease(35).<br>• Due to aging, there is loss of functioning nephrons in the kidney. Differences in aging between sexes are recorded but mechanisms poorly understood. | |
| **Pregnancy and lactation(20, 36, 37):** | |
| • Due to the additional workload for the fetus, there are alterations in glomerular filtration rate, electrolyte and balance during pregnancy. | |
| **Menstrual cycle(38, 39):** | |
| • The increase and decrease of estrogen and progesterone levels during the different phases of the menstrual cycle can lead to change in blood flow and | |
| fluid balance, retention of more sodium and loss of iron. | |

The origin of sexual dimorphism in humans can be traced back to the allocation of sex chromosomes at conception. The presence of a Y chromosome typically results in the development of male genitalia and the production of male hormones, while the absence of a Y chromosome results in female genitalia and hormone production. Throughout development and life, sexual dimorphism is also influenced by differences in gene expression, epigenetics, sex hormones, and environmental factors (see **Table 1).** Epigenetic mechanisms such as X-chromosome inactivation and imprinting, cause gene expression differences in the developing fetus depending on its sex. Sex steroids influence gene expression through sex-steroid receptors on the cell surface, resulting in permanent epigenetic changes(20). Levels of sex-steroid fluctuate throughout different life stages, with events such as mini-puberty and puberty marking drastic rises in hormone levels, and hormone levels declining for both sexes around the age of 50, see **Figure 2**. Specifically, for men, the primary sex hormone testosterone remains stable around 4.7 ng/ml post-puberty. The hormone system in women, however, is far more complex. After puberty, females experience a 28-day cycle of progesterone, luteinizing hormone (LH), estradiol, and follicle-stimulating hormone (FSH) **(Figure 3**), with estradiol levels fluctuating between 30 and 800 pg/ml during the cycle(40). Lifestyle, aging, and pregnancy then further exaggerate the differences.

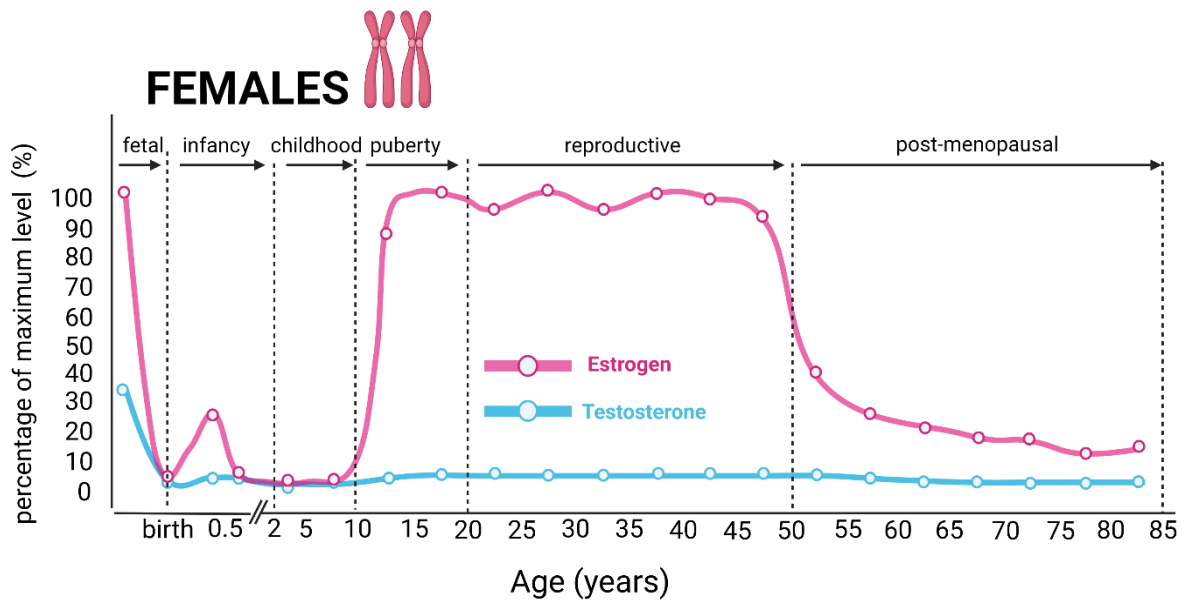

**Figure 2:** Approximate testosterone and estradiol levels in the plasma of men and women from before birth up to the age of 80. Levels are shown as a percentage of the maximum mean of testosterone and estradiol respectively. The figure is adapted from(41).

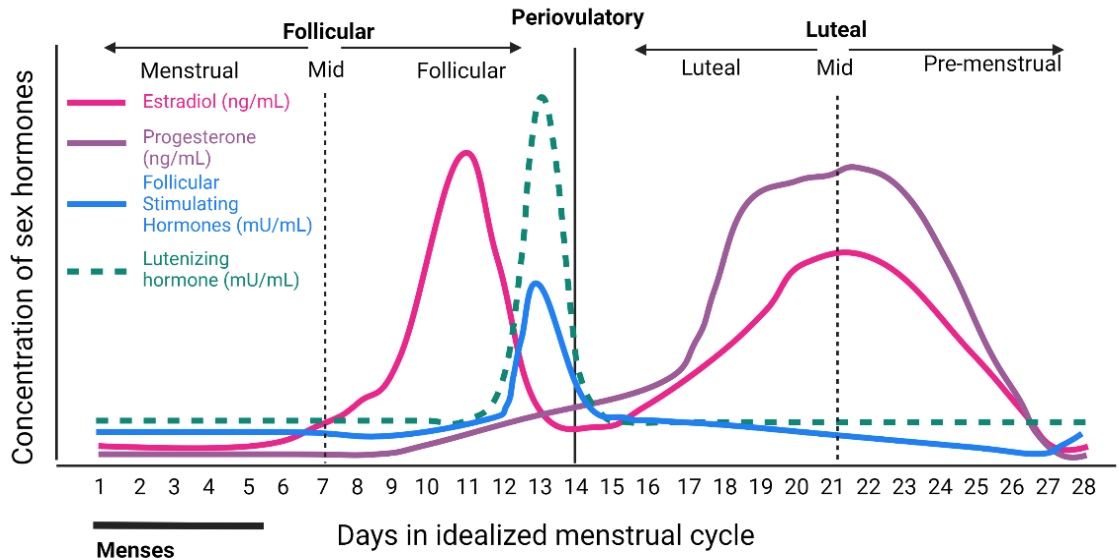

**Figure 3:** Idealized female hormone cycle of 28 days. Changing levels of estradiol (ng/ml), progesterone (ng/ml), luteinizing hormone (mU/ml), and follicular stimulating hormone (mU/ml) are shown. Adapted from(42).

## 3 Sexual Dimorphism of the kidney

Sexual dimorphism in the kidney refers to the physical and functional differences between male and female kidneys, although both sexes have similar kidney anatomy (with respect to body surface area (43)). One notable difference is the weight of the kidney relative to body weight, which is smaller in female than in males. Additionally, female kidneys have a lower total nephron count (16). In fact, in female children, glomeruli - the part of the kidney responsible for blood filtration- are significantly larger than in age-matched male children (44). These differences in filtration rate and size mean that cross-validation models used to estimate glomerular filtration rate (GFR) whether creatinine- or cystatin C-based, often include sex-specific factors for all age groups (45, 46). Below the major functional sex-dimorphisms of the kidney are elaborated.

### 3.1 Transporter machinery

Transporters are the proteins present on the membrane of the cells that facilitate diffusion, absorption and secretion of different solutes in/out of the lumen resulting in managing the urine concentration. The pattern of transporters along the nephron is sexually dimorphic. Veiras *et al.* found that female rat proximal tubuli have higher levels of Na$^+$/H$^+$ exchanger 3 (NHE3) and lower levels of Na$^+$-Pi cotransporter 2, aquaporin-1 (AQP1), and claudin-2 compared to the male rats nephrons. In contrast, the distal tubuli of female rats has higher levels of Na$^+$-Cl$^-$ cotransporter (NCC), claudin-7, and epithelial Na⁺ channel (ENaC) α- and γ-subunits (47). These transporters are highly specific and efficient and their coordinated activity is important in determining sex dependent solute handling. A modelling study found that to induce urine excretions identical to those seen in the male rat nephron model, a variety of transporter activities in the female rat nephron model must be upregulated or downregulated in certain segments of the nephron. Within that model, fewer solutes and filtered volume are reabsorbed in the female proximal tubule compared to male rats. Conversely, in female rats a greater proportion of Na+ and water transport occurs in the distal tubule and the thick ascending limb, overall leading to a similar urinary output (48). Sex-specific patterns of transporters in the kidney persist even under a high-sodium diet, with females expected to exhibit 40% lower NHE3 and 200% higher NCC abundance (49).

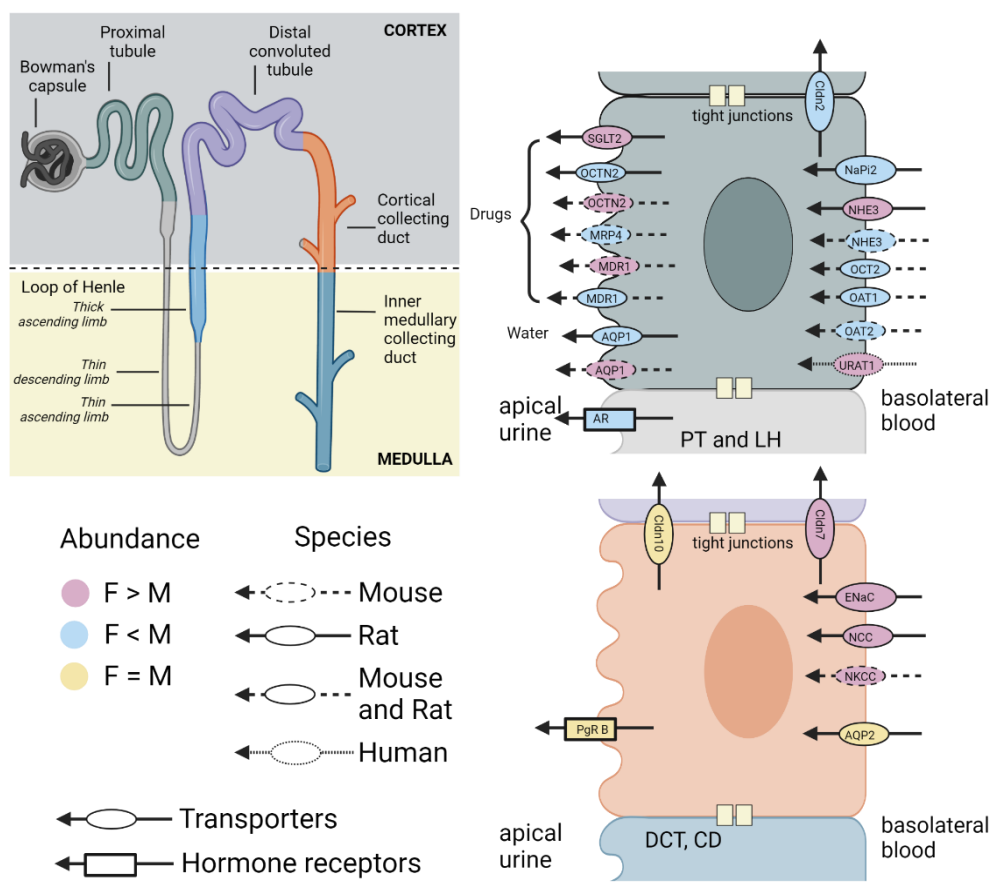

**Figure 4:** Main transporter and receptor activity in the proximal tubule cells of the nephron, the most important site of drug and endogenous metabolites exchange in the kidney and distal tubule cells. Sex-based differences in activity of these transporters has been reported by sex, species and abundance.

**Table 2**: Protein abundance by species and sex.

| Species | Proteins | Abundance | Reference |
|---|---|---|---|
| Mouse | SLC17A3, SLC3A1, SLC22A28, SLC1A6 | M > F | (50, 51) |
| | SLC39A5, SLC6A19, SLC7A7, SLC7A8, SLC3A2 | F > M | |
| Rats (pregnant and lactating) | ENaC ↑, Na-K-ATPase ↓ | - | (52) |
| Rats (pregnant) | Urinary angiotensin (1-7) ↑, AQP1 ↑ | - | (53) |
| Rats | MDR1, OAT1, OAT2, OCTN2 | M > F | (54) |
| | SGLT2 | F > M | |
| Mouse | MDR1, OCTN2 | F > M | |
| | OAT2, MRP4 | M > F | |
| Dog | MDR1 | F > M | |

Hormones like estrogen, progesterone and prolactin stimulate NCC expression resulting in a higher NCC activity in female rats (55). Similar abundance in NCC activity has also been seen in mice (56) and could very well be translated to humans. An overview of transporter activity is shown in **Table 2** and the complexity of (sex-specific) transport mechanisms along the nephron is given in **Figure 4**.

Thus, pharmacokinetic, safety, efficacy, and toxicity data acquired from a non-human test model, might drastically misrepresent sex-based mechanisms in humans, so caution must be exercised when extrapolating such data to the sex/species not included in the experiment (54).

## 3.2 RAAS system

The regulation of arterial pressure and renal function under RAAS exhibits a sex bias, with males showing upregulation of the gene expression of renin, angiotensin-II and aldosterone. This upregulation leads to a more evident rise in blood pressure in men than in women upon exposure to angiotensin II. In contrast, the female sex exhibits enhanced expression of the counter-regulatory arm of the RAAS system, including angiotensin-covering enzyme (ACE) 2/Ang(1-7)/MasR and AT2R which may contribute to the lower prevalence of hypertension and cardiovascular disease in pre-menopausal women compared to men (57, 58).

Estrogen and testosterone levels play a significant role in the control of blood pressure through the RAAS system (3). Estrogen increases angiotensinogen levels, but negatively affects renin, ACE, AT1 receptor, and aldosterone. Additionally, it inhibits the activation of NAD(P)H by ACE2, which reduces the production of reactive oxygen species and helps maintain vascular function. In contrast, progesterone competes with aldosterone for mineralocorticoid receptors, which can lead to reduced salt and water retention and lower blood pressure. In turn, testosterone raises plasma levels of renin and ACE, which may lead to an increased risk of hypertension in men.

Pre-menopausal women experience protection against high blood pressure, renal, and cardiovascular diseases, which is lost with menopause and the associated loss of female sex hormone load (3, 57). This loss is mainly due to the decrease in estrogen levels, which lead to increased vascular tone, inflammation, and oxidative stress. Therefore, estrogen replacement therapy with may be beneficial for post-menopausal women suffering of hypertension or cardiovascular disease, although this needs to be balanced against the potential risks associated with hormone therapy, such as increased risks of ovarian and breast cancer, thromboembolic disease, etc (59). Overall, understanding the sex differences in RAAS regulation and the effects of sex hormones on blood pressure control is important for the development of personalized and effective treatment strategies for hypertension and cardiovascular disease, but also a proper management of such conditions.

### 3.3 Sex hormone receptors in the kidney

Hormone receptors play a crucial role in regulating the activity of the hormones, which in turn control the signal cascading. The influence of sex steroids on the kidney has been well-documented, and research as early as 1998 has shown that testosterone can affect the expression of renal transporters (NHE3, NKCC2, AQP2) (60). Estrogen and androgen have been found to stimulate calcium membrane transport in tubule cells *in vitro (61)*, while progesterone has been identified as a high-affinity ligand for the mineralocorticoid receptor, which controls electrolyte transport in the kidney (62).

Localizing receptors for sex hormones in the mammalian kidney has been complicated by the limited abundance of the receptor proteins and/or their mRNAs, and by the complexity of the relationship between hormone receptors and specific actions within the kidney, which remains an area of ongoing research.

| Sex | Receptor type | Localization | F vs M | Reference |
|---|---|---|---|---|
| M | ERα, ERβ | Interstitial cells, collecting duct | - | (63) |
| F & M | ERβ | Cortex > Medulla, Convoluted PT | - | |

| Sex | Receptor type | Localization | F vs M | Reference |
|---|---|---|---|---|
| F & M | PgR B | Medulla > Cortex, Glomerulus, distal tubule, interstitium (nuclei) | F = M | |
| F & M | ARα > ARβ | - | F = M | |
| F & M | AR | - | F < M | |
| F & M | AR | Distal tubule | - | |
| F &M | AR | - | F < M | (64, 65) |
| F &M | AR | - | F = M | (64) |

**Table 3** gives an overview of the localization of known estrogen, androgen, and progesterone receptors in the kidney, the sex of the cells they were detected in, and if sex-specific abundance differences were recorded (63).

The consequences of alterations in hormone receptor activity can be significant, as disruptions to the normal functioning of the kidney can lead to a range of clinical conditions. For example, changes in the activity of mineralocorticoid receptors can result in disorders such as hypertension and electrolyte imbalances. Additionally, alterations in the expression of renal transporters can lead to impaired kidney function, which can contribute to the development of chronic kidney disease and other related conditions. Understanding the role of hormone receptors in the kidney is therefore an important area of research, as it has the potential to inform the development of new treatments and therapies for a range of kidney-related conditions. Ongoing studies in this field will continue to shed light on the complex interactions between hormones, hormone receptors, and kidney function, and will help to guide the development of effective interventions to improve patient outcomes.

| Sex | Receptor type | Localization | F vs M | Reference |
|---|---|---|---|---|
| M | ERα, ERβ | Interstitial cells, collecting duct | - | (63) |
| F & M | ERβ | Cortex > Medulla, Convoluted PT | - | |
| F & M | PgR B | Medulla > Cortex, Glomerulus, distal tubule, interstitium (nuclei) | F = M | |
| F & M | ARα > ARβ | - | F = M | |
| F & M | AR | - | F < M | |
| F & M | AR | Distal tubule | - | |
| F &M | AR | - | F < M | (64, 65) |
| F &M | AR | - | F = M | (64) |

**Table 3:** Sex hormone receptors in the human kidney. Receptor type, localization within the kidney, and relative distribution between males and females are marked. Male sex is denoted as M, female sex as F. Estrogen receptors alpha and beta, progesterone receptors, and androgen receptors are denoted as ER, PgR, and AR respectively.

## 3.4 Nephrotoxicity

Nephrotoxicity, the toxic side effects of chemicals on the kidneys, can have significant clinical consequences. The loss of nephrons, the functional units of the kidneys, can lead to impaired kidney function, and in severe cases, kidney failure. The kidney is particularly vulnerable to drug-induced toxicity because of its high metabolic activity, the concentrations of solutes in its tubules and its role in drug clearance(66). Nephrotoxic compounds can harm all parts of the kidney, including specific cell types, like podocytes and proximal tubule epithelial cells, and disrupt regulatory mechanisms, like glomerular filtration rate or the RAAS system(67).

As discussed above in 3.1, sex differences in drug transporters expression can results in differences in drug clearance rates and the nephrotoxicity of drugs associated with that transporter. For example, women have a 25% lower systemic clearance of the anticoagulant lepirudin by the kidney than men (68) and are more likely to suffer nephrotoxicity induced by the antibiotic amikacin (69). Peri-menopausal women are at significantly higher risk for cisplatin-induced nephrotoxicity than men, an effect that is currently explained by the higher estrogen level in women (10).

The late detection of drug-induced nephrotoxicity during drug development process is a significant problem. Animal models are not always predictive of nephrotoxicity in humans because of interspecies variations, which further complicates the process (70). Therefore, there is a need for reliable and early detection models for drug-induced nephrotoxicity models that take into consideration sex-dependent pharmacokinetics and toxicities profiles. .

## 3.5 Kidney Diseases

Differences between men and women have been recorded for many kidney-related diseases. CKD is estimated to affect over 10 % of the global population(71). In its contribution to end-stage kidney disease (ESKD) and cardiovascular disease, CKD represents a global morbidity and mortality burden(72). CKD shows sex-related differences in pathophysiology, symptoms, complications, and therapeutic efficacy. As such women, have a higher incidence of CKD, but men tend to progress towards ESKD at a faster rate. This difference could be due to the effect of sex hormones on nitric oxide metabolism, which has renoprotective properties. Estrogen, in particular, is thought to have a major role in the slower progression of CKD in women, as once in menopause, a severe decline is noted (13, 73).

In diabetes, women are more likely to develop renal complications, but do so with an approximately 10-year delay compared to their male counterparts. Sex hormones and sex-specific gene expression are thought to be at fault(74). They have shown that the olfactory receptor, Olfr1393 gene has a stronger phenotype in female rats than male. The authors theorize that the difference in glucose/glutamine metabolism may explain the sex differences of kidney disease in diabetes(12). Other studies that have also recorded genes that are overexpressed or underexpressed in the subjects with renal diseases and related to pregnancy, spermatogenesis and other sex-related differences are recorded in **Table 4**.

**Table 4:** Genes associated with various kidney diseases that shows sex dependence. Genes related to pregnancy and spermatogenesis are highlighted in bold.

| Gene | Associated disease | Reference |
|---|---|---|
| JAG1, HES1 | Diabetic nephropathy (human) | (75) |
| GAP43, FOXD1, POSTN, EMP2, **PTGIS**, MPPED2, **AKR1C2**, **PTHLH**, ANKRD6, LRRC17, **VIM** | Chronic kidney disease (human) | (76) |
| CA9, DGCR5, EGLN3, SLC16A3, SLC5A3, SPAG4, VEGF, SEMA5B, PFKP, PRAME, MUF1, MGC45419, GPR54, FABP7 | Clear cell renal cell carcinoma (human) | (77, 78) |
| Tpbg, RFP, HLA-DQA1, SCNN1B, SLC15A2 | X-linked Alport syndrome (Canine) | (26) |
| VEGFa, LRPAP1, AGTR1, SLC12A1, CUBN, DAO1, KCNJ1, CASR, PTHR1, SLC12A3, FXYD2, KLK6, SCNN1A, SCNN1B, ATP6V1B1, AVPR2, AQP2 | Polycystic kidney disease (mouse) | (79) |

Men with polycystic kidney disease have a worse prognosis compared to women. This can partially be explained by the effect of testosterone on cAMP generation. Testosterone increases cAMP generation in kidney cells, which stimulates fluid secretion and solute transport (80). Traditionally, men are at higher risk of developing kidney stones, but disease prevalence has been rising in the last decade, especially for women(14).

X-linked chromosome inactivation in females could lead to Alport syndrome, X-linked dominant hypophosphatemic rickets, X-linked nephropathy and Fabry disease (23, 25). Female children showed more prevalence of lupus nephritis than their male counterparts, however, the disease has been reported to be more aggressive in male adults(81, 82). Pregnancy related kidney diseases such as pre-eclampsia and gestational diabetes only occur in females. Pre-eclampsia females showed an increase sodium reabsorption which could result in hypertension(83). Because of these clinical differences, it is important to consider sex differences in studies of kidney disease, as otherwise important mechanisms may be overlooked.

## 4 Engineered kidney models

The development of accurate kidney models is crucial for advancing our understanding of kidney function and disease, as well as for testing the safety and efficacy of new therapies. As mentioned earlier, there are significant sex differences in the kidney, so advanced kidney models must enable the recapitulation of both male and female kidney phenotypes. Therefore, it is important to develop kidney models that incorporate these sex-specific differences in order to better understand the mechanisms underlying these differences and to evaluate potential therapeutic interventions. This can be achieved through the use of animal models, such as mice, that have been genetically engineered to reflect human physiology and disease, or through the use of *in vitro* models, such as organoids, that can be generated from human stem cells and cultured in a dish. By using these models, researchers can gain a better understanding of how sex-specific genes, hormones, and physiological differences contribute to kidney function and disease, and identify potential targets for therapy. In summary, accurate kidney models that incorporate sex-specific differences are critical for advancing our understanding of kidney disease and developing effective treatments.

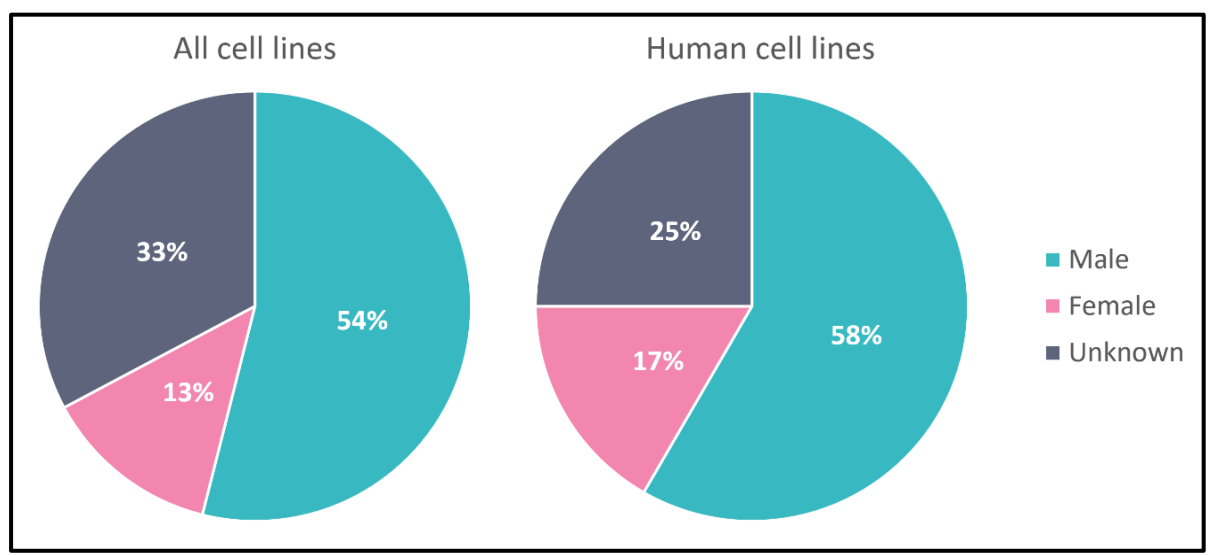

**Figure 5:** Recorded sex of 128 kidney cell lines for all species (left) found in the cell line database Cellosaurus(84) and specifically 42 human cell lines (right). Cell lines are separated

by recorded sex, male or female. Cells for which no sex is recorded are denoted as unknown. Detailed information can be found in the appendix.

## 4.1 Cell sources

Sex differences in kidney function and injury response change with age(9, 85). The extent to which this relationship carries over to *in vitro* experiments is unclear. Age- and sex-specific changes in DNA methylation indicate that these factors likely cannot be considered separately(86). Kouthouridis *et al.* found that up to 50% of *in vitro* studies do not specify the sex of their cells and only 20% use cells of both sexes (40).

*In vitro* kidney modelling is achieved through the use of primary cells, immortalized cell lines, and induced pluripotent stem cells (iPSCs). Primary cells closely mimic *in vivo* physiology, but are limited by the time and labour-intensive isolation process, and their culture is restricted by dedifferentiation and senescence. Immortalized cell lines have been developed to overcome the limitations of primary cells, but most experience some degree of dedifferentiation resulting in low predictivity or transporter expression loss. It should be remarked that in human kidney disease research, non-human donor species are more frequent than human cells (87, 88). The sex of the donor is not always made known in primary sources for cell lines due to patient confidentiality and ethics guidelines. Instead, for established cell lines, the sex of the donor is most of the time noted in the initial studies, but in the follow ups the information is lost. To illustrate both the distribution of male and female sex and the number of cell lines with unknown sex, we had a closer look at the database Cellosaurus (84). We identified an unequal distribution of sexes within established cell lines, with a third of recorded cell lines lacking information about their sex (**Figure 5**+**Appendix A**). For a quarter of the human kidney cell lines, the sex could be determined and two-thirds of the remaining cell lines were male.

Sex distribution varied by nephron segment (**Table 6**) with the sex of all human glomerular cell lines remaining unknown. For the thick ascending loop of Henle, there are neither human cell lines nor cell lines with known female sex. Only one embryonic kidney cell line was noted to be female, no male comparison could be found in the database.

**Table 6:** Original nephron segment/cell type and sex of recorded cell lines in **Appendix A**. One kidney pair represents one cell line. Male cell lines are indicated in blue, and female cell lines in pink. Cell lines with no determined sex are indicated in grey.

|  | Non-human cell lines | Human cell lines |
|---|---|---|
| Collecting duct | (many) | (few) |
| Distal convoluted tubule | (few) | |
| Distal tubule | (few) | |
| Embryonic kidney | | (one) |
| Epithelial-like | (few) | (one) |
| Glomerular | (few) | (few) |
| Kidney / Diverse | (few) | |
| Mesangial | (few) | (few) |
| Podocyte | (one) | (few) |
| Proximal convoluted tubule | (few) | (few) |
| Proximal tubule | (many) | (many) |
| Renal cell carcinoma | (one) | (many) |
| Thick ascending limb of loop of Henle | (few) | |

immortalized cell lines develop chromosomal instability after repeated passages. For several cell lines a loss of the Y chromosome has been recorded, inhibiting their use in sex-specific studies (89).

Besides primary cells and immortalized cell lines, the use of induced pluripotent stem cells presents a third alternative. Using iPSC technology, a variety of kidney cell types and kidney stem cells can be created from patient material. Cells are isolated and through exposure to a cocktail of factors reprogrammed back to pluripotency. The iPSCs can then be redifferentiated towards most cell types. In the paper of Ribeiro *et al.,* protocols for differentiating iPSCs into various kidney cell types are collected. However, the use of iPSCs is not without its drawbacks. In human pluripotent cells, the expression of the sex-determining Y-chromosome-linked gene SRY in males cell leads to differential autosomal gene expression compared to female cells. This could affect steroid metabolism and differentiation (90). Furthermore, donor sex influences the nature of the imprinting defects in iPSCs and their response to culture conditions (91). The creation of iPSC-derived kidney cells is time-consuming and resource-intensive(92, 93). As with primary cells, the availability of knowledge about iPSC donor sex depends on patient confidentiality and ethics guidelines.

## 4.2 2D Models

Although 2D cell culture are still the most commonly used model for drug screening and nephrotoxicity studies (94) due to their simplicity and practicality, they have limitations in terms of physiological relevance (87, 95). These limitations include the lack of apical-basal polarization and expression of key transporters (67, 87), as well as lack of cell-ECM interactions which can affect cell proliferation, polarization, migration, signal transduction, and gene expression (96). This translates to low accuracy of 2D models in predicting nephrotoxicity. Additionally, the proximal tubule is often the focus of nephrotoxicity testing. A systematic review by Irvine *et al.* found that of the included *in vitro* nephrotoxicity studies, 82% studied proximal tubule cells (94). Although they are particularly susceptible, this fails to capture the variety of nephrotoxicity, which can also affect all parts of the kidney (66). The development of *in vitro* kidney models which capture the simplicity, practicality, and speed of 2D models, while delivering translatable results, has therefore long been strived for.

## 4.3  3D Models

Due to the complexity of the kidney and the presence of different cell types, using single cell-type 2D assays to study kidney function of toxicity may not fully capture the physiological interactions and complexities of the organ. This could lead to misinterpretations of the data and the potential for important effects to be overlooked. This could happen both when testing for nephrotoxicity but also when creating kidney models to answer other questions. Therefore, the complexity of a model must match the research question. In recent years, a multitude of more complex kidney models have been developed.

### 4.3.1  Organoids

Organoids are a 3D culture system in which stem cells self-organize *in vitro* to form patterns highly similar or identical to their native organs. Kidney organoids can be generated both from iPSCs and adult stem cells (AdSCs). iPSC-derived organoids are established by guiding iPSCs through the developmental steps of the wanted cell population. AdSC organoids are established through the isolation of the cells from the adult tissue, they are maintained for organoid culture through the presence of the niche factors specific to this stem cell type (93, 97, 98). AdSC organoids are therefore restricted to tissues with maintained stem cell niches (99). Organoids present an accessible, rapid, robust, and genetically stable human model system that forms physiologically relevant 3D structures. It allows for patient-specific creation and genetic manipulation and is accompanied by fewer ethical concerns than animal models. On the other hand, organoids are a pricey culture method that lacks standardization and homogeneity between different individuals, assays, and groups. This leads to often heterogenous outcomes between publications (93). Higher-order kidney organoids include cells from the nephron progenitor lineage and the ureteric bud lineage. These organoids can display the full architecture of the embryonic kidney including differentiated nephrons connected to a ureteric epithelium (100).

Kidney organoids can be used to study kidney development, disease, drug safety and as a cell source for regenerative therapies. Especially the option of genetically modifying organoids has allowed for the better study of kidney diseases caused by genetic mutations, for example, autosomal dominant polycystic kidney disease. As these organoids can be compared to their isogenic negative controls, causalities can be explored more clearly than in the previously available mouse or patient-derived cell models (88).

First attempts have been made to use kidney organoids in nephrotoxicity studies. Although organoids with multiple different kidney cell types have been developed, markers are missing which are sensitive enough to distinguish toxicity in the different cell types. Organoids are also difficult to perfuse, limiting compound administration (88, 96).

Although not yet reported for kidneys, organoids have been used to study the impact of sexual dimorphisms in other organs. Kelava *et al.* studied the impact of sex chromosomes and sex steroids on neurogenic potential by creating brain organoids from stem cells of both sexes and the addition of androgens and estradiol to the culture medium (101). Devall *et al.* compared male and female organoids in a study on the influence of calcium in the context of colorectal cancer (102). These studies emphasize the future impact of organoids as a platform for the study of sex dimorphism and its impact on potential therapies.

### 4.3.2 Kidney-on-a-chip

Organs-on-a-chip are microfluidic culture devices that enable cell culture in a highly regulated environment. This culture method is defined by its capacity to provide continuous perfusion, the incorporation of sensors, physical stimulation, and see-through materials that allow for visual observation using microscopy. The architecture of a chip is highly customizable but often consists of multiple perfusable channels separated by a membrane. The characteristics of the used plastics can be replaced by hydrogels to mimic natural ECMs. The channels within the construct can then be seeded with single or multiple cell types (103). For close recapitulation of physiology the correct structures, cell-cell interactions, electrochemical and osmotic pressure gradients, flow dynamics, as well as cellular metabolic and endocrine functions must be established (104).

Since the first kidney-on-a-chip was established by Jang *et al.* in 2010 several different models for different parts of the nephron have been developed (105). Among them were models for the glomerulus, distal tubule, and collecting duct (104). The most developed kidney-on-a-chip models are developed for the proximal tubule. Here, microphysical models already closely reflect the physiology of the renal proximal tubule, including basolateral solute transport, apical solute uptake, and intracellular enzymatic function. Within these models, several different cell types were integrated and toxic injury response could be modelled (106). Schutgens *et al.* demonstrated that by combining patient-derived AdSC organoids and microfluidic systems, a platform could be built that allowed for molecular and cellular analysis, disease modelling, and drug screening while being rapid and personalized. This method could also be utilized using stem cells obtained from cystic fibrosis patients' urine, implying that in the future patient-specific, non-invasive disease-on-a-chip models are possible (99). The glomerulus was modelled by layering human podocytes and glomerular endothelial cells. These chips exhibit glomerular functions such as permeability and selectivity and can be maintained in long-term culture (107).

Kidneys-on-chips have been offered as an optimal platform for drug nephrotoxicity testing. Their setup makes the cells more accessible than organoids while maintaining physiological relevance, such as cell polarity and function (104, 108). An on-chip model of vascularized kidney spheroids within integrated sensors enabled rapid elucidation of nephrotoxicity mechanisms of Cyclosporine and Cisplatin (109).

Well-characterized *in vitro* models of the thick ascending limb, interstitial, and cortical collecting duct are yet to be developed. Additionally, the finely tuned interactions of the individual nephron parts make the development of a whole kidney on a chip more than the connections of singular models. But ongoing work in the co-culture of epithelial and endothelial cells as well as advancements in creating gradients chip platforms have paved the way for more complex and relevant on-chip models (104).

Because organ-on-chip systems are highly customizable and regulatable, they offer unique opportunities to incorporate and study sex differences. A review of organ-on-chip studies by Nawroth *et al.* found that organ-on-chip models are uniquely qualified to study the effect of soluble factors such as hormones, as levels can be varied continuously to mimic the physiological concentration within a tissue (110). Although a variety of male and female reproductive organs have been modelled in organ-on-chip systems, consideration of sex as a biological variable for other organs is lacking (110, 111). The review by Nawroth *et al.* found that only 3% of the considered organ-on-chip platforms recorded the use of solely female cells within one model system (110). No studies were found that used the distinctive benefits of organ-on-chip platforms, such as physical stimulation, sensors, and channel geometry, to study sexual dimorphism. Therefore the potential of organ-on-chip technology in exploring sex differences in kidneys and nephrotoxicity testing has yet to be fully realized.

### 4.3.3 Biofabricated kidneys

By combining additive manufacturing techniques with biocompatible materials and cells, bioprinting allows for the creation of complex biological structures. A variety of bioinks and biomaterial inks have been developed and used in the context of kidney bioprinting, such as hydrogels made up of decellularized ECM. For the creation of kidney models, primarily extrusion-, microfluidic- and droplet-based printers have been used (112). Especially interesting is the use of coaxial nozzles and sacrificial ink, which allows for the creation of hollow tubes with distinct inner and outer layers for the recreation of nephron structures (113).

Bioprinting is often used in combination with other technologies. Through the combination of microfluidic and bioprinting techniques, geometrically complex models can be made. One example is the development of a perfuse-able convoluted renal proximal tubule with modifiable geometries (114). Extrusion bioprinting has also been used to form kidney organoids in a highly uniform and high-throughput manner, enabling their use in nephrotoxicity testing (115). Organovo Inc. has developed a 3D-printed tissue specialized for nephrotoxicity testing. Within that model an interstitial layer, a basement membrane, and a polarized layer of renal epithelium are cultured on top of each other, possibly giving insight into cell-cell interactions during nephrotoxicity (116). These applications highlight the possible benefits of bioprinting kidneys. It allows for the creation of highly customizable shapes and stratifications in a reproducible, high-throughput manner. Although to this date no studies on sex-specific biofabrication have been published, these benefits of bioprinting will be vital in the future.

### 4.3.4 De-& Recellularized Kidneys

Kidney ECM is highly specific to each segment of the nephron in its composition and architecture. It supports tissue integrity and stimulates kidney cells but is also integral to kidney function and carefully maintained by surrounding cells. One example is the glomerular basement membrane, an integral part of the glomerular filtration barrier, which is formed and cultivated by a combination of podocytes, endothelial cells, and mesangial cells (117). Corresponding to its importance, some attempts at engineering kidney models have focused on preserving and reseeding the ECM of cadaver kidneys.

Detergent treatment can remove cells and cell debris from the matrix without damaging the structure and composition. Glomerular and tubular structures remain intact, preserving the functions mediated by the ECM. The intact structure of the blood vessels simplifies the perfusion of freshly bioengineered kidneys. The cell removal process also removes HLA antigens. Thus, recellularized kidneys combined with autologous cells offer a possibility for immunosuppressive free transplantation (118, 119). Because of the different cell types within the kidney and its intricate structure, recellularization is complex. By seeding endothelial cells through the vasculature, epithelial cells through the collecting system, and utilizing a trans renal pressure gradient, a cell site specific repopulation of a rat kidney could be achieved. Recellularized kidneys produced urine *in vitro* and after orthotopic transplantation *in vivo*, were shown to clear metabolites *in vivo*. Despite this recellularized kidneys remained immature due to incomplete seeding, immaturity of seeded cells, and decreased graft perfusion. Upscaling towards human application is hindered by the expansion of the necessary cells and development of the de- and recellularization techniques for human-sized organs (119).

There are few papers recording ECM composition or structure differences between the male and female sex and none could be found for the kidney. But existing literature for bone, brain, and cancer indicates that ECM composition in health and disease is sexually dimorphic (120-122). In the future, de- and recellularization of healthy and diseased (male and female) ECM could aid in the functional recapitulation of the specific tissue phenotypes in the laboratory.

## 4.4 *In silico* models

Next to 2D and 3D models of kidney development, great strides have been made in the development of renal *in silico* models. Mathematical models of the whole kidney and singular nephrons have been constructed, helping to unravel the roles of renal transporters (123, 124). For example, Weinstein et al. developed spatial mathematical models of tubular transport to calculate cytosolic concentrations, transport fluxes and epithelial permeabilities (124-127). *In silico* models can also be used to predict nephrotoxicity, but predictions made so far are generalized in nature and do not, for example, give information on the type of adverse effects or toxic dosage (128, 129). Hallow et al. created a whole kidney compartmental model to determine time response of single nephron GFR, sodium and glucose balance, tubuloglomerular feedback under varying conditions (130-132). Edwards et al. made a protein uptake model along the proximal tubule to explain proteinuria (133). Three-dimensional rendering of kidney vasculature from CT-scans of rat kidney were proposed by Nordsletten et al. (134). In an exciting approach of combining the vascular architecture of the inner medullary with an epithelial transport model of Layton et al., Pannabecker et al. tried to explain the urine concentrating mechanism thereof (135). Notably, *in silico* models have always been a handoff between the mathematical models themselves and the experimental data they are founded on. Because more information on rat kidneys is available, most current models focus on rat kidneys. Similarly, the above models do not include sex differences, mostly due to lack of detailed experimental data. As such, of big importance for the topic of this review are the mathematical models developed by the Layton group. In recent years, they have published several mathematical models of sex differences in rat kidneys, such as transporter pattern differences and their functional implications as well as differences in circadian rhythm regulation and hypertension (11, 48, 136, 137). Their *in silico* model of the effect of pregnancy on the female rat kidney revealed that renal transporter sex differences may be in preparation for the strain of pregnancy (138). These models have helped us understand how sex differences translate to differences in kidney function and have further highlighted the need for sex representation in biomedical research, including sex specific experimental data for *in silico* model calibration and validation.

## 5 Design criteria for sex-specific in vitro kidney models

Developing kidney models tailored to sexspecific differences within engineered human kidney tissues presents numerous benefits for exploring disparities in renal (patho)physiology. To start, these humanbased models sidestep the constraints linked to species-specific effects frequently encountered in animal studies. The incorporation of iPSC)technology allows for the implementation of patient-specificity, ensuring an essentially boundless supply of patient-specific cells that can be expanded and directed into various cell types, including renal lineages. Furthermore, dissecting the roles of sex-related factors in clinical and preclinical investigations often confronts obstacles posed by confounding variables. The utilization of sex-specific tissues provides a precise means of managing environmental and behavioral factors, enabling a focused exploration of genetic, epigenetic, and hormonal variables. Lastly, these models offer a platform for meticulous fine-tuning of experimental variables. This includes the ability to precisely control factors like the timing and dosage of hormone (pre)conditioning. The manipulation of the cellular microenvironment becomes feasible through adjustments in the stiffness, degradation, and composition of biomaterials, closely mirroring the properties of the kidney matrix under both healthy and pathological conditions. Additionally, gene editing techniques can be effectively employed to selectively modify specific genes, either to create disease models or to rectify genetic anomalies. Collectively, these advantages empower researchers to conduct quantitative, mechanistic inquiries into sex-specific distinctions in renal physiology in vitro. This presents a valuable opportunity to deepen our comprehension of kidney health and disease within the context of sex, fostering a more comprehensive understanding of these complex factors.

Incorporating sexual dimorphism into engineered kidney models offers multiple key benefits. Sexual dimorphism is species-specific, limiting the applicability of data from animal studies(94, 139). Data from human models could therefore provide better insights into human sex-specific physiology. *In vitro* models offer a highly controlled environment where experimental settings and complexity can be adjusted according to the experimental question. This way the influence of genetic, epigenetic, and hormonal differences between the two sexes can be studied for the kidney within a highly controlled environment. Sexually dimorphic models for nephrotoxicity could make clinical testing safer for female participants as women (especially pregnant women) are more likely than men to suffer adverse drug reactions (94, 140). More accurate preclinical safety and efficacy testing will also save money during the drug development process as up to 19% of drugs currently fail during phase III clinical testing due to nephrotoxicity issues not detected previously (70).

Not every experiment needs to examine sex differences. However, the sex of the biological materialst used should always be noted. This increases the reproducibility of the experiments and prevents results from being incorrectly extrapolated to the other sex. However, it would be prudent to critically examine and discuss the sex-specificity needed for each study to adequately reflect human physiology (4).

## 5.1 Cells

Selecting cells is often the first step in creating an *in vitro* model. It should therefore also be the first step of sex incorporation.

- Record donor sex of the cell lines. If not provided by vendors, use of PCR to determine sex should be considered.
- Report loss of Y chromose after repeated cell passages.
- Journals could mandate the reporting of sex when cell lines are used (141).

Thus, the first step towards incorporating sex-dimorphism in *in vitro* models is the documentation of the sex of the cells used, regardless of cell line, primary cell, or iPSCs. Whenever possible, cells of both sexes should be incorporated into the study and the sex of the cells evaluated as a biological variable (**Figure 6**).

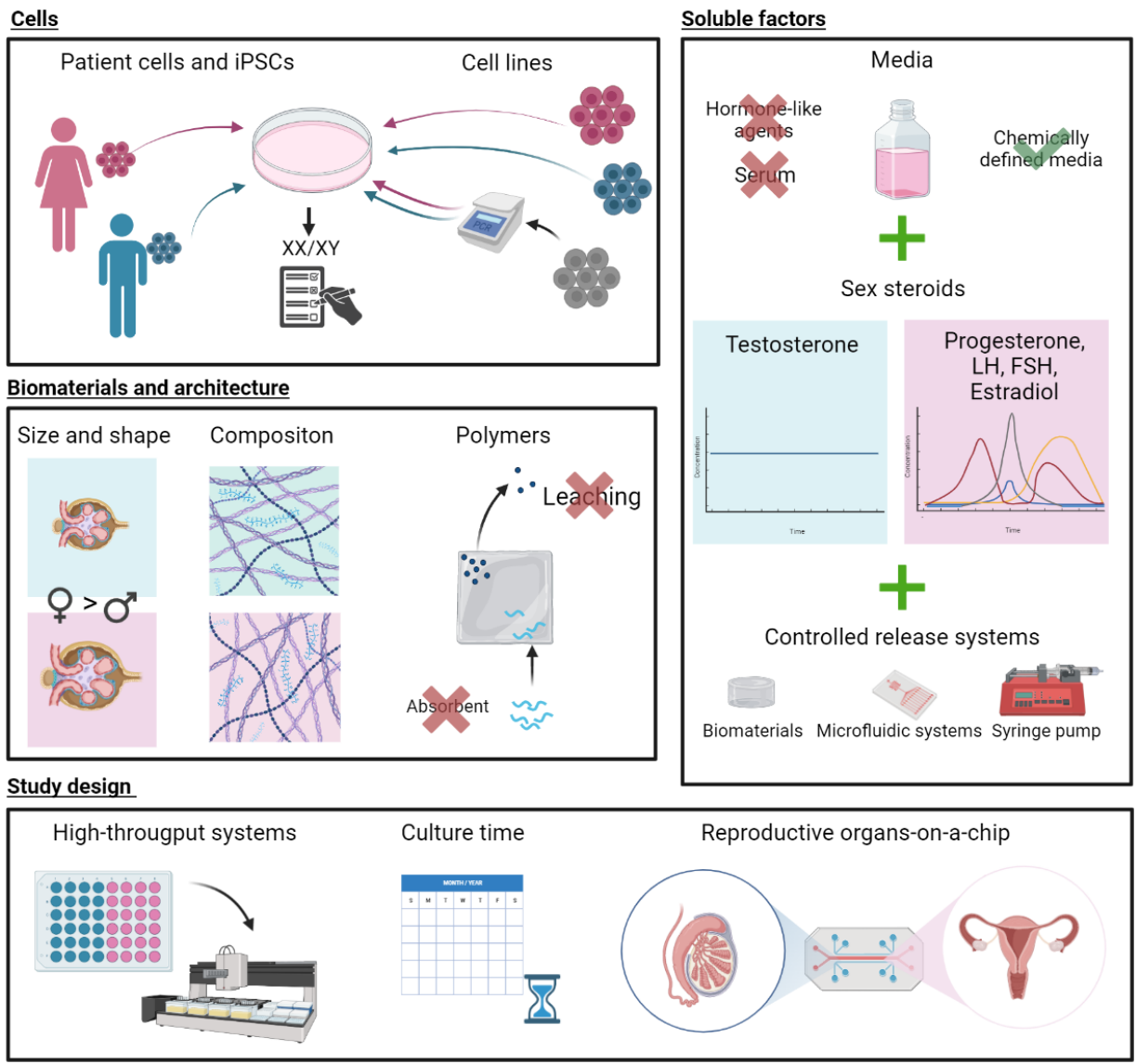

**Figure 6**: Opportunities to incorporate sex differences *in vitro* models through cells, soluble factors, biomaterials, and architecture as well as study design. Male sex is indicated with blue, female sex with pink. Undetermined sex is indicated in grey.

## 5.2 Soluble factors

Incorporating sex hormones into culture media is another opportunity to impart sexual dimorphism within model systems. Sex-steroid levels vary considerably during different life stages as illustrated in **5.** Sex hormone levels in adult women additionally fluctuate in an approximately 28-day cycle, see **Figure** . Accordingly, while the addition of testosterone to the medium might be enough to model an adult male, more complex models will be needed for the female sex to capture different phases of the menstrual cycle, the changes in the kidney function during pregnancy and lactation (40). Furthermore, serum hormone levels are not equivalent to tissue levels because the endothelial barrier is responsible for both concentration and temporal differences (142). To avoid overdosing, specific tissue levels should therefore be carefully evaluated. Effects recorded under supraphysiologic conditions may not reflect reality.

Advancements for controlled drug delivery could be adapted for the controlled delivery of sex hormones to *in vivo* models. Functionalized biomaterials may both passively or upon stimulation release their load (143). Similarly, microfluidic devices developed for controlled *in vivo* drug delivery may be utilized and offer even more control over hormone levels (144). Optimal would be the inclusion of the hormone-producing gonads within human-on-chip models. Promising examples are a female reproductive tract-on-chip model which replicated the 28-day menstrual cycle (145) and the multi-organ chip co-culture of liver and testis cells (111).

Not only the addition of sex-specific hormones to the media is important, but also the elimination of all other hormones or hormone-like agents that could confound the experiment. Serum is a frequent addition to the medium. Sera not only contain sex hormones (146), but analysis has shown that concentrations of sex hormones may change from batch to batch, potentially causing variation in experimental outcome (147). The often-used method of stripping cell culture media through the use of charcoal filters to remove sex hormones, may further increase batch to batch variability and generally decreases serum efficacy (148). Serum-free media formulations offer a chemically defined and consistent alternative to traditional animal-derived media. But serum-free alternatives to traditional culture sera are difficult to develop because of serum complexity. Culture conditions were traditionally developed by trial and error, but the degree of complexity involved in serum optimization makes this approach too time intensive. Kim and Audet developed a serum-free media formulation for hematopoietic cells using a high-dimensional search algorithm. This shows one way the development of serum-free media alternatives could be optimized (149, 150). Another benefit is that the use of serum-free media has been associated with decreased inter-donor variability (149). Using chemically defined media also removes medium batches as a variable between experiments.

Unwanted sex hormone-like agents in media also need to be considered. Sex hormone receptors are present on kidney cells, as elaborated in

| Sex | Receptor type | Localization | F vs M | Reference |
|---|---|---|---|---|
| M | ERα, ERβ | Interstitial cells, collecting duct | - | (63) |
| F & M | ERβ | Cortex > Medulla, Convoluted PT | - | |

| F & M | PgR B | Medulla > Cortex, Glomerulus, distal tubule, interstitium (nuclei) | F = M | |
|---|---|---|---|---|
| F & M | ARα > ARβ | - | F = M | |
| F & M | AR | - | F < M | |
| F & M | AR | Distal tubule | - | |
| F &M | AR | - | F < M | (64, 65) |
| F &M | AR | - | F = M | (64) |

**Table** , and a variety of non-hormone compounds can bind and activate those receptors (151). Phenol red, included in most standard media as a pH indicator, is a known estrogenic compound. It has been shown to affect cell proliferation rates, stem cell differentiation, and the efficacy of some anti-cancer drugs. Male and female cells respond differently to exposure to phenol-red. To further complicate matters, concentrations of phenol red differ between media types and suppliers (40, 148). To exclude the effect of hormone-like agents on study outcome, all medium additions should be carefully considered for potential hormone activity (**Figure 6**).

## 5.3 Materials and architecture

Sexual dimorphism also needs to be considered when choosing the materials and the architecture of the models. ECM composition differences between males and females have been shown for several tissues such as the brain and heart (121, 152). Such information about the kidney is lacking and this knowledge gap should be closed. Nevertheless, it is essential for correct modelling to reconstruct potential differences *in vitro*. Synthetic polymers can be adjusted in mechanical and chemical characteristics. When using natural polymers, the sex of the source should be recorded and if necessary, the sex of the ECM source and model should be matched (141).

When designing 3D micro-physiological models, the sex differences in kidney architecture can also be considered. Glomerular size is sex-dependent both in human children and mice (44, 153). 3D printing offers the opportunity to incorporate such differences within models.

It is critical that materials used for microfluidic devices are biocompatible, inert, and non-leaching. When using microfluidic devices in combination with carefully controlled hormone levels, it is important to consider all involved materials (**Figure 6**). Polydimethylsiloxane (PDMS), a frequently used polymer for microfluidic devices, absorbs hydrophobic molecules and may thus change the concentration of added hormones. For the construction of biochips, materials that exhibit less absorption of hydrophobic compounds, like polycarbonate and poly(methyl methacrylate) should be favoured (104). The materials used in cell culture could also be a source of hormone-like agents. The polystyrene used for most single-use cell culture plastics has been shown to release weak estrogens and may weaken the effects of added estrogens (148).

## 5.4 Study design

The study design is severely affected by the inclusion of sex as a biological variable. The simple inclusion of models for both sexes doubles the sample size. Should sex be considered a variable and tested for, the sample size could multiply by six (40). This highlights the importance of high throughput systems for the incorporation of sex as a biological variable. Shaughnessey *et al.* have developed a high throughput system to evaluate kidney toxicity, exemplifying ways in which multiple independent samples can be assessed simultaneously (154).

The long female cycle additionally complicates the study setup and brings up several questions. When comparing a female model system with a full 28-day hormone cycle and a male system on stable testosterone, specific days of the cycle should be chosen for direct comparison and not deviated from within that study (89). Because hormone levels can affect physiology, the days of the female cycle most optimal for testing should be explored. It is also not yet known how long cells must be conditioned within a hormone cycle to fully adapt to it. If this period is too long, the integration of the female cycle into cell culture experiments could be too complex for many setups. Lastly, going through the entire cycle would prolong *in vitro* studies dramatically. This makes it worthwhile to explore hormone concentrations and configurations of hormones which allow for an approximation of the menstrual cycle effect.

As discussed in section 3, kidney function changes with age in a sex-dependent manner, including drug-dependent sensitivity (9). For high-fidelity kidney models of all potential patients, different stages of human life and the according sex-specific kidney function may need to be represented (**Figure 6**).

First steps have been taken to create sex-specific toxicity models. In the case of Baert *et al.* a co-culture of liver and testis organoids was set up to test for the reproductive toxicity of a drug metabolite from the liver. This highlights both the importance of incorporating multi-organ systems and sex-specificity in toxicity tests (111). A design focusing on sex-specific engineered cardiac model was proposed by Lock et al. (22).

## 6 Discussion

Sex-specificity needs to be included in design of kidney models, in vitro and in vivo. This enables targeted development of drugs including both sexes. Models can also help gap the understanding between the drug efficiency between different sexes. Effect of the menstrual cycle, pregnancy and lactation could be looked at greater detail when we ensure sex-inclusivity in research.

The need for sex-inclusive studies in kidney research is empashized by the presence of sex hormone receptors and their effect of sex hormones on kidney function, as seen in the RAAS, Therefore, the development of human sex-specific models for drug testing is of utmost importance. Better *in vitro* and *in silico* model systems will enable the study of individual contributing factors.

Novel techniques such as organoids, kidney-on-a-chip, biofabrication, and recellularization offer excellent opportunities for sex-specific modelling. However, these opportunities remain underutilized, and most studies using these models do not account for sex-specific differences. The potential value of sex-inclusive studies of the kidney has been exemplified by Clotet-Freixas *et al.* and Sandhu *et al.* (12, 80). Because these studies recapitulated the sex differences in kidney disease, they were able to give insights into previously unknown disease mechanisms.

Despite the potential value of sex-inclusive studies of the kidney, there are major challenges ahead in incorporating sex dimorphism in engineered kidney models. For example, the availability of kidney cell lines specific to each part of the nephron for each sex is severely limited. A comprehensive library of cell lines with known sex and a wide demographic range would enable easier modelling. Another challenge is the culture method. In particular, truly sex-neutral culture methods need to be created (as a first step as well as control). They should be equally conducive to the culture of male and female cells, and the medium and serum used should include no hormones or hormone-like agents. Leaching or absorbent materials should be avoided. To such a culture method, sex-specific signals like hormones can then be added. Down the line, controlled release systems will enable the incorporation of more advanced hormone fluctuations to mimic e.g. the menstrual cycle. Importantly, it is not yet clear to what degree physiological circumstances need to be imitated to recreate sex-specificity. While the use of cells of both sexes and sex hormones might be enough in some cases, a more intricate setup might be needed for others. A deeper understanding of the origin of sex dimorphisms and their evolutionary purpose in the kidney could provide the needed insight.

The development of advanced, sex-specific *in silico* models also faces significant hurdles. One of the major challenges is the lack of available data on sex-specific differences in renal physiology and disease. Comprehensive studies exploring transporter differences, the quantity and location of sex hormone receptors, and ECM composition and structure in the male and female human kidney are missing. The current data heavily relies on animal models, but the interspecies variation calls the applicability of such data into question. As the quality of *in silico* models relies heavily on the data of previous studies, better experimental data from human models will also help the development of accurate human *in silico* models of the kidney.

While this review focused on the kidney, most of the opportunities for sex-specificity *in vitro* are equally applicable to other tissues of the body. As the importance of sex differences in health and disease becomes more well-known, the methods to include them in *in vitro* models will become more common and refined. This will enable a better understanding of the underlying mechanisms of disease throughout human physiology, overall leading to more effective treatments that account for sex-specific differences.

# 7 Conclusions

In conclusion, the inclusion of sex-specific differences in engineered *in vitro* and *in silico* kidney models is vital for advancing our understanding of kidney physiology and disease. It is also essential for developing more accurate and predictive drug development and safety assays. To this end, a framework has been proposed for creating sex-specific kidney models (*in vitro* and *in silico*), which includes cell selection and characterization, the surrounding extracellular environment (structure and hormones) and appropriate study design.

# 8 Declarations

***Ethics approval and consent to participate***
Not applicable

***Consent for publication***
Not applicable

***Availability of data and materials***
Not applicable

***Competing interests***
The authors declare that they have no competing interests.

***Funding***
This work is supported by the partners of Regenerative Medicine Crossing Borders (RegMed XB), a public-private partnership that uses regenerative medicine strategies to cure common chronic diseases, by the Dutch Kidney Foundation and Dutch Ministry of Economic Affairs by means of the PPP Allowance made available by the Top Sector Life Sciences & Health to stimulate public-private partnerships (DKF project code PPS08) and by a grant from the Dutch Kidney Foundation (22OK1018).

## Authors' contributions

CV, SM and SS participated in writing the manuscript, preparing the figures and tables. SM, SS, AC have all reviewed and edited the manuscript. All authors read and approved the final manuscript.

## Acknowledgements

Not applicable

# 9  Appendix A

Renal cell lines in the database Cellosaurus(84). Cell lines are sorted alphabetically first by donor species, then by name. Parent cell lines are listed in bold, and popular cell line children are noted when applicable. Information about donor sex, species, and cell or tissue origin is added whenever available. Cells lines with known use in nephrotoxicity testing are highlighted in light grey.

| Name | Species | Cell type | Sex |
| --- | --- | --- | --- |
| **NH-GEN1** | Bovine | Glomerular endothelial | Unknown |
| **MDCK** | Dog | Distal tubule and collecting duct | Female |
| **A6** | Frog (African clawed) | Distal tubule and collecting duct | Male |
| **BHK** BHK-21 | Hamster (Syrian) | Kidney | Unknown |
| **769-P** | Human | Renal cell carcinoma | Female |
| **786-O** | Human | Renal cell carcinoma | Male |
| **A-498** | Human | Renal cell carcinoma | Male |
| **A-704** | Human | Renal cell carcinoma | Male |
| **ACHN** | Human | Papillary renal cell carcinoma | Male |
| **AP3** | Human | Cortical collecting duct | Male |
| **AP8** | Human | Cortical collecting duct | Female |
| **BB64RCC** | Human | Renal cell carcinoma | Male |
| **BFTC-909** | Human | Renal pelvis urothelial carcinoma | Male |
| **C2M12/16** | Human | Mesangial | Unknown |
| **CAKI-2** | Human | Papillary renal carcinoma | Male |
| **CAL-54** | Human | Renal cell carcinoma | Male |
| **CiGEnC** | Human | Glomerular endothelial | Unknown |
| **CIHGM-1** | Human | Mesangial | Male |
| **ciPEC** | Human | Parietal epithelial | Unknown |
| **ciPTEC** | Human | Proximal tubule epithelial | Female |
| **HA7-RCC** | Human | Renal cell carcinoma | Male |
| **HCD** | Human | Cortical collecting duct | Unknown |
| **HEK293** | Human | Embryonic kidney | Female |
| **HGVEC.SV1A4** | Human | Glomerular epithelial cell | Unknown |
| **HK-2** HKC | Human | Proximal tubule | Male |
| **HKC-5;8;11** | Human | Proximal tubule epithelial | Unknown |
| **hPCT-03-ts** | Human | Proximal convoluted tubule | Female |
| **hPCT-05-wt** | Human | Proximal convoluted tubule | Male |
| **hPCT-06-wt** | Human | Proximal convolute tubule | Male |
| **HREC24T** | Human | Proximal tubule epithelial | Male |
| **HRTPT** | Human | Proximal tubule epithelial | Male |
| **KMRC-1** | Human | Clear cell renal carcinoma | Male |
| **KMRC-2** | Human | Clear cell renal carcinoma | Male |
| **KMRC-20** | Human | Clear cell renal carcinoma | Unknown |
| **KMRC-3** | Human | Clear cell renal carcinoma | Male |
| **KTCTL-13/RCCER** | Human | Clear cell renal carcinoma | Male |
| **KTCTL-140** | Human | Clear cell renal carcinoma | Female |
| **KTCTL-195** | Human | Clear cell renal carcinoma | Male |
| **KTCTL-1M** | Human | Clear cell renal carcinoma | Male |
| **KTCTL-21** | Human | Clear cell renal carcinoma | Male |
| **KTCTL-26A** | Human | Clear cell renal carcinoma | Male |
| **LB1047-RCC** | Human | Renal cell carcinoma | Female |
| **LB2241-RCC** | Human | Renal cell carcinoma | Male |
| **LB996-RCC** | Human | Renal cell carcinoma | Male |
| **NCC010** | Human | Renal cell carcinoma | Male |
| **NCC021** | Human | Renal cell carcinoma | Unknown |
| **OS-RC-2** | Human | Clear cell renal carcinoma | Male |
| **PODO/TERT255** | Human | Podocyte | Male |
| **PODO/TERT256** | Human | Podocyte | Female |
| **RCC10RGB** | Human | Renal cell carcinoma | Male |
| **RCC4** | Human | Clear renal cell carcinoma | Unknown |
| **RPTECs TERT/1** | Human | proximal tubule | Male |
| **RS** | Human | Proximal tubule epithelial | Male |
| **RXF 393L** | Human | Renal cell carcinoma | Male |
| **SA7K** | Human | Proximal tubule epithelial | Female |
| **SLR-20** | Human | Renal cell carcinoma | Female |
| **SLR-21** | Human | Clear cell renal carcinoma | Unknown |
| **SLR-23** | Human | Clear cell renal carcinoma | Unknown |
| **SLR-25** | Human | Clear cell renal carcinoma | Unknown |
| **SLR-26** | Human | Clear cell renal carcinoma | Male |
| **SN12C** | Human | Renal cell carcinoma | Male |
| **SNU-1272** | Human | Clear cell renal carcinoma | Female |
| **SNU-349** | Human | Clear cell renal carcinoma | Male |
| **SV-HGEC** | Human | Glomerular endothelial | Unknown |
| **SW156** | Human | Renal cell carcinoma | Male |
| **TH1 /TH7** | Human | Proximal tubule epithelial | Unknown |
| **TK-10** | Human | Clear renal cell carcinoma | Male |
| **TUHR10TKB** | Human | Clear cell renal carcinoma | Male |
| **TUHR14TKB** | Human | Renal cell carcinoma | Male |
| **UM-RC-2** | Human | Clear cell renal carcinoma | Unknown |
| **UM-RC-6** | Human | Renal cell carcinoma | Male |
| **UO-31** | Human | Renal cell carcinoma | Female |
| **UOK101** | Human | Clear cell renal carcinoma | Unknown |
| **VMRC-RCW** | Human | Renal cell carcinoma | Male |
| **VMRC-RCZ** | Human | Renal cell carcinoma | Unknown |
| **CV-1** COS-1/3/7 | Monkey (African green) | Kidney | Male |
| **VERO** | Monkey (African green) | Kidney epithelial | Female |
| **JTC-12** | Monkey (Cynomolgus) | Renal proximal epithelium | Female |
| **209/MDCT** | Mouse | Distal convoluted tubule | Male |
| **CWRU-C57BL** | Mouse | Proximal tubule epithelial cells | Unknown |
| **DKC-8** | Mouse | Distal convoluted tubule epithelial | Unknown |
| **KDT3** | Mouse | Distal tubule | Unknown |
| **KPT1** | Mouse | Proximal tubule | Unknown |
| **KPT2** | Mouse | Proximal tubule | Unknown |
| **M-1** | Mouse | Cortical collecting duct | Unknown |
| **mCCDcl1** | Mouse | Collecting duct principal cell | Unknown |
| **MCT** | Mouse | Proximal tubule | Unknown |
| **mCT1** | Mouse | Cortical collecting duct epithelial | Unknown |
| **mIMCD-3** | Mouse | Inner medullary collecting duct | Unknown |
| **mIMCD-K2** | Mouse | Inner medullary collecting duct | Male |
| **mPEC** | Mouse | Parietal epithelial | Male |
| **mpkCCd** | Mouse | Cortical collecting duct | Unknown |
| **mpkDCT** | Mouse | Distal convoluted tubule | Unknown |
| **mpkIMCD** | Mouse | Inner medullary collecting duct | Unknown |
| **MuRTE** | Mouse | Proximal tubule epithelial | Male |
| **MuRTE61** | Mouse | Cortex | Male |
| **PKSV-PCT** | Mouse | Proximal convoluted tubule | Male |
| **PKSV-PR** | Mouse | Proximal straight tubule | Male |
| **ST-1** | Mouse | Thick ascending limb of loop of Henle | Unknown |
| **SV40-tsA58** | Mouse | Proximal convoluted tubule | Unknown |
| **TKC2 /TKD2** | Mouse | Tubule epithelial | Unknown |
| **TKPTS** | Mouse | Proximal tubule epithelial | Unknown |
| **OK** | Opossum (American) | Epithelial-like | Female |
| **LLC-PK1** | Pig (Hampshire) | Epithelial-like | Male |
| **Clone C** | Rabbit | Renal intercalated cell | Unknown |
| **CNT** | Rabbit | Distal tubule | Male |
| **PAP-HT25** | Rabbit | Inner medullary epithelium | Unknown |
| **PST-S2** | Rabbit | Proximal tubule epithelial cell | Male |
| **RPT-I8** | Rabbit | Proximal tubule epithelial | Male |
| **TALH SVE 1** | Rabbit | Thick ascending limb of loop of Henle | Unknown |
| **vEPT** | Rabbit | Proximal tubule epithelial | Male |
| **CCD-IC** | Rabbit (New Zeeland white) | Cortical collecting duct | Male |
| **RC.SV** | Rabbit (New Zeeland white) | Proximal tubule | Male |
| **RCCT-28A** | Rabbit (New Zeeland white) | Cortical collecting duct | Unknown |
| **2DNA1D7** | Rat | Podocyte | Male |
| **HBZY-1** | Rat | Mesangial | Unknown |
| **IRMC** | Rat | Mesangial | Male |
| **RGE** | Rat | Glomerular endothelial | Female |
| **rRCCd5** | Rat | Renal tubule cell carcinoma | Male |
| **SGE-1** | Rat | Glomerular epithelial | Male |
| **NRK** NRK-52E | Rat (Norway) | Proximal tubule | Unknown |
| **SKPT-0193** | Rat (spontaneously hypertensive) | Proximal tubule | Male |
| **RCCD1** | Rat (Sprague Dawley) | Cortical collecting duct | Male |
| **RCCD2** | Rat (Sprague Dawley) | Cortical collecting duct | Male |
| **raTAL** | Rat (Sprague-Dawley) | Medullary thick ascending loop of Henle | Male |
| **SGE-1** | Rat (Wistar) | Glomerulus | Male |
| **WKPT-0293** | Rat (Wistar) | Proximal tubule | Male |

| WKPT-1292 | Rat (Wistar) | Proximal tubule | Male |
|---|---|---|---|

**Nomenclature**

    CA9: Carbonic anhydrase 9

    DGCR5: DiGeorge syndrome critical region gene 5

    EGLN3: Egl-9 family hypoxia-inducible factor 3

    SLC16A3: Solute carrier family 16 member 3

    SLC5A3: Solute carrier family 5 member 3

    SPAG4: Sperm-associated antigen 4

    VEGF: Vascular endothelial growth factor A

    SEMA5B: Semaphorin 5B

    PFKP: Phosphofructokinase platelet

    PRAME: Preferentially expressed antigen in melanoma

    MUF1: Mitochondrial ubiquitin fold modifier 1

    MGC45419: Uncharacterized protein C14orf166

    GPR54: G protein-coupled receptor 54

    HIG2: Hypoxia-inducible protein 2

    Tpbg: Trophoblast glycoprotein

    RFP: Regenerating islet-derived family, member 4

    HLA-DQA1: Major histocompatibility complex, class II, DQ alpha 1

    SCNN1B: Sodium channel, non-voltage-gated 1 beta subunit

    SLC15A2: Solute carrier family 15 member 2

    JAG1: Notch ligand jagged 1

    HES1: Hairy and enhancer of split 1

    FADD: Fas-associated protein with death domain

    DAXX: Death domain-associated protein

    p53: Tumor protein p53

    BAD: Bcl-2-associated agonist of cell death

    PITX1: Paired-like homeodomain transcription factor 1

    HOXC10: Homeobox protein Hox-C10

    ARHGAP26: Rho GTPase-activating protein 26

    TACSTD2: Tumor-associated calcium signal transducer 2

    GAP43: Growth-associated protein 43

    FOXD1: Forkhead box D1

    POSTN: Periostin

    EMP2: Epithelial membrane protein 2

    PTGIS: Prostaglandin I2 (prostacyclin) synthase

    MPPED2: Metallophosphoesterase domain-containing protein 2

AKR1C2: Aldo-keto reductase family 1 member C2

PTHLH: Parathyroid hormone-like hormone

ANKRD6: Ankyrin repeat domain-containing protein 6

LRRC17: Leucine-rich repeat-containing protein 17

VIM: Vimentin

VEGFa - Vascular Endothelial Growth Factor A

LRPAP1 - Low Density Lipoprotein Receptor-Related Protein Associated Protein 1

AGTR1 - Angiotensin II Receptor Type 1

SLC12A1 - Solute Carrier Family 12 Member 1

CUBN - Cubilin

DAO1 - D-Amino Acid Oxidase 1

KCNJ1 - Potassium Inwardly Rectifying Channel Subfamily J Member 1

CASR - Calcium Sensing Receptor

PTHR1 - Parathyroid Hormone Receptor 1

SLC12A3 - Solute Carrier Family 12 Member 3

FXYD2 - FXYD Domain-Containing Ion Transport Regulator 2

KLK6 - Kallikrein-Related Peptidase 6

SCNN1A - Sodium Channel Epithelial 1 Alpha Subunit

SCNN1B - Sodium Channel Epithelial 1 Beta Subunit

ATP6V1B1 - ATPase H+ Transporting V1 Subunit B1

AVPR2 - Arginine Vasopressin Receptor 2

AQP2 - Aquaporin 2

Olfr1393 - Olfactory receptor family 139 subfamily 3 member 93

> Recommended readings:
> 1. James BD, Guerin P, Allen JB. Let's Talk About Sex—Biological Sex Is Underreported in Biomaterial Studies.. *Adv. Healthcare Mater.* 10, 2001034 (2020). https://doi.org/10.1002/adhm.202001034
> 2. Lock R, Al Asafen H, Fleischer S, Tamargo M, Zhao Y, Radisic M, Vunjak-Novakovic G. A framework for developing sex-specific engineered heart models. Nat Rev Mater 7, 295–313 (2022). https://doi.org/10.1038/s41578-021-00381-1
> 3. Pardue ML, Wizemann TM, editors. Exploring the biological contributions to human health: does sex matter? (2001).